\documentclass{aa}
\usepackage{graphicx}
\usepackage{txfonts}
\usepackage[breaklinks, colorlinks, citecolor=blue, urlcolor = blue]{hyperref}
\usepackage{amsmath}
\usepackage{amssymb}
\usepackage{url}
\usepackage{graphicx}
\usepackage[normalem]{ulem}
\usepackage[utf8]{inputenc}
\usepackage{multicol}
\usepackage{multirow}
\usepackage{adjustbox}
\usepackage{mathtools}
\usepackage{longtable}
\usepackage{pdflscape}
\usepackage{color}
\usepackage{soul,xcolor}
\setstcolor{purple}

\newcommand{\order}[1]{} %to impose the correct order in reference with same first author and same year
\newcommand \sw{{\it Swift}}

\newcommand \fe{{\it Fermi}}

\begin{document}

\title{The rise and fall of the high-energy afterglow emission of GRB 180720B} %choose a title
\titlerunning{Afterglow of GRB 180720B}  

\author{
M. Ronchi\inst{1}\thanks{E--mail: m.ronchi26@campus.unimib.it}
\and F. Fumagalli\inst{1}
\and M. E. Ravasio\inst{1}
\and G. Oganesyan\inst{2, 3}
\and M. Toffano\inst{4, 5}
\and O. S. Salafia\inst{5, 6}
\and L. Nava\inst{5, 7, 8}
\and \\ 
S. Ascenzi\inst{5}  
\and G.~Ghirlanda\inst{5,6} 
\and G. Ghisellini\inst{5}  
}
%put in right order and with right affiliations (some might miss)

\authorrunning{M. Ronchi et al.}
%G. Ghirlanda, O. S. Salafia, G. Ghisellini, G. Oganesyan, M. E. Ravasio, M. Ronchi, F. Fumagalli and M. Toffano}

\institute{
$^1$ Università degli Studi di Milano-Bicocca, Dip. di Fisica ``G. Occhialini'', Piazza della Scienza 3, I-20126 Milano, \\
$^2$ Gran Sasso Science Institute, Viale F. Crispi 7, I-67100, L’Aquila (AQ), Italy\\
$^3$ INFN - Laboratori Nazionali del Gran Sasso, I-67100, L’Aquila (AQ), Italy\\
$^4$ Università degli Studi dell'Insubria, Via Valleggio 11, 22100, Como, Italy\\
$^5$ INAF -- Osservatorio Astronomico di Brera, Via E. Bianchi 46, I-23807 Merate, Italy\\
$^6$ INFN -- Sezione di Milano-Bicocca, Piazza della Scienza 3, I-20126 Milano, Italy \\
$^7$ INFN -- Sezione di Trieste, via Valerio 2, I-34149 Trieste, Italy \\
$^8$ Institute for Fundamental Physics of the Universe (IFPU), I-34151 Trieste, Italy
}

\date{Received xxx / Accepted: xxx}

\abstract{
The Gamma Ray Burst (GRB) 180720B is one of the brightest events detected by the \fe\ satellite and the first GRB detected by the H.E.S.S. telescope above 100\,GeV. 
We analyse the \fe\ (GBM and LAT) and \sw\ (XRT and BAT) data and describe the evolution of the burst spectral energy distribution in the 0.5\,keV--10\,GeV energy range over the first 500 seconds of emission. 
We reveal a smooth transition from the prompt phase, dominated by synchrotron emission in a moderately fast cooling regime, to the afterglow phase whose emission has been observed from the radio to the GeV energy range. 
The LAT (0.1--100\,GeV) light curve initially rises ($F_{\rm LAT}\propto t^{2.4}$), peaks at $\sim$78\,s, and falls steeply ($F_{\rm LAT}\propto t^{-2.2}$) afterwards. 
The peak, which we interpret as the onset of the fireball deceleration, allows us to estimate  the bulk Lorentz factor $\Gamma_{0}\sim 150 \ (300)$ under the assumption of a wind-like (homogeneous) circum-burst medium density.
We derive a flux upper limit in the LAT energy range at the time of H.E.S.S. detection, but this does not allow us to unveil the nature of the high energy component observed by H.E.S.S.
We fit the prompt spectrum with a physical model of synchrotron emission from a non-thermal population of electrons. The 0--35\,s spectrum after its $E F(E)$ peak (at 1--2 MeV) is a steep power law extending to hundreds of MeV. We derive a steep slope of the injected electron energy distribution $N(\gamma)\propto \gamma^{-5}$. 
Our fit parameters point towards a very low magnetic field ($B'\sim 1 $\,G) in the emission region.}
  
 %add others

\maketitle

\section{Introduction}

Despite having been discovered more than fifty years ago, from a theoretical point of view gamma-ray bursts (GRBs) still represent one of the most challenging transient sources of high-energy photons in the Universe. 
Even if some of the observational features can be explained within the context of the standard fireball model developed by \cite{Rees1992,Meszaros1993,Rees1994}, there are still many important missing pieces of the GRB complex puzzle (see \citealt{Meszaros2019, Zhang2019, Kumar2015, Piran1999, Piran2004}). 
The physical origin of the prompt emission, for example, is still under debate.
Although the synchrotron process has been proposed as the emission mechanism responsible of the observed radiation \citep{Rees1994, Sari1996, Sari1998}, for a long time the predictions of the spectral shape from a population of accelerated electrons cooling via the synchrotron process have been thought to be inconsistent with the shape of the typical observed GRB prompt emission spectra \citep{Preece1998, Ghisellini2000, Kaneko2006, Ghirlanda2009, Nava2011, Gruber2014}. 
For many years, this inconsistency has been the major argument against the synchrotron interpretation of the GRB prompt emission spectra. 
Only recently, new observational evidences were found in support of the synchrotron interpretation \citep{Oganesyan2017,Oganesyan2018,Oganesyan2019, Ravasio2018, Ravasio2019}. 
Extending the investigations down to X-ray and optical bands, a remarkable consistency of the observed GRB prompt spectra and synchrotron predictions has been discovered when electron cooling is taken into account.
\cite{Burgess2018} also showed that most of the time-resolved spectra of single-pulse \fe\ GRBs can be successfully fitted by the synchrotron model.

The extension of the energy range towards higher energies (> 100 MeV) played a fundamental role in the understanding of the nature of the high energy component produced by these powerful transient sources. The Large Area Telescope (LAT) \citep{Atwood2009} on board the \fe\ satellite detected and characterized the high energy emission of several GRBs \citep{Ajello2019}. The long lasting high energy (100 MeV-100 GeV) emission has been interpreted (see \citealt{Nava2018} for a review) as due to the external shock afterglow of the burst \citep{Kumar2009, Kumar2010, Ghisellini2010, Beniamini2015}. However, the characterization of the transition from the prompt to the afterglow dominance is hampered by evidences of a superposition of these two emission components early on in the burst. 

GRB 190114C, the first GRB detected by the MAGIC telescopes \citep{Mirzoyan2019} with high significance, opened a new window on the interpretation of the high and very high energy emission (VHE, above 100\,GeV). 
The analysis of the spectrum at lower energies showed a clear example of the mixture of the prompt and the afterglow components in the same energy band. 
Using \fe\ Gamma-Ray Burst monitor (GBM) + Large Area Telescope (LAT) 
and the Neil Gehrels Swift Observatory (hereafter \sw)  X-Ray telescope (XRT) + Burst Alert Telescope (BAT) data, \citet{Ravasio2019b} have shown the rise of an additional non-thermal component in the keV-MeV energy range in the very early phases ($\sim 5$\,s after trigger time) of the prompt emission. 
This component, fitted by a power law $dN/dE\propto E^{\Gamma_{\rm PL}}$ with slope $\Gamma_{\rm PL} \sim -2$,  was interpreted as the synchrotron emission from the afterglow of the burst. The difficulties in interpreting the VHE emission of GRB190114C as synchrotron from relativistic electrons \citep{Guilbert1983}, led to ascribe its origin to synchrotron self-compton (SSC) \citep{Derishev2019, Fraija2019, Ravasio2019b, Wang2019}.

Recently, GRB 180720B was claimed to be detected at VHE by the H.E.S.S. telescope \citep{CTASymposium}. This GRB is also one of the brightest bursts ever detected by the \fe\ satellite, with a fluence $f= 2.99 \cdot 10^{-4}$ erg  cm$^{-2}$ in the energy range 10 - 1000\,keV \citep{Roberts2018}. Its keV--MeV prompt emission spectrum has been analysed, together with other 9 long bright GRBs, in \citet{Ravasio2019} who showed the consistency of the spectral shape with synchrotron emission. It has also been detected by LAT and reported in the LAT GRB catalogue of \citet{Ajello2019}. Therefore, GRB180720B is another interesting event for studying its spectral energy distribution (SED) extending from keV to GeV energies and possibly unveil the transition from the prompt to the afterglow dominance and constrain the physical parameters of the emission process.

This paper is based on the study of the evolution of the spectrum of GRB~180720B during the first 500\,s after trigger time, from an analysis of the data of \fe\ (LAT+GBM) and \sw\ (BAT+XRT) satellite. In particular, we extract the light curves and analyse the spectra in order to study the temporal evolution of the spectral energy distribution and the consistency with the synchrotron emission, shown in \S \ref{sec:results}. Interpreting the peak of the LAT light curve as due to the onset of the deceleration of the jet, we derive an estimate of the bulk Lorentz factor $\Gamma_0$ in the cases of a uniform density profile and of a stellar wind density profile for the circum-burst medium (\S \ref{sec:gamma0}). 
We discuss the theoretical implications of our results and summarise the conclusions in \S \ref{sec:discussion} and \S \ref{sec:conclusions}.

\section{Data reduction and analysis}\label{sec:data_analysis}

On 20 July 2018 at 14:21:39 UT, \fe/GBM triggered GRB 180720B \citep{Roberts2018}, which was also detected by \sw/BAT \citep{GCN_Swift} and \fe/LAT \citep{GCN_LAT}. The burst was also detected by Konus-Wind \citep{Frederiks2018}, by the CALET Gamma-Ray Burst Monitor \citep{GCN_CALET} and in the X-ray band by MAXI/GSC \citep{GCN_MAXI} and by NuSTAR \citep{GCN_NuSTAR}. GRB~180720B was observed by an intensive follow-up campaign. \sw/XRT began observing the source 86.5 seconds after the BAT trigger. \cite{GCNottico} detected a bright optical counterpart, 73 seconds after the trigger using the 1.5-m Kanata telescope. They measured an R-band apparent magnitude of $m\sim 9.4$ which corresponds to an optical flux $E F(E) \sim 1.4$ keV cm$^{-2}$ s$^{-1}$  at a central frequency of about $4 \times 10^{14}$ Hz. 
Several other optical ground telescopes observed this burst \citep{GCNottico22979, GCNottico22983, GCNottico22985, GCNottico22988, GCNottico23004, GCNottico23017, GCNottico23020, GCNottico23021, GCNottico23023, GCNottico23024, GCNottico23033, GCNottico23040}.
The redshift was measured with the instrument X-shooter on VLT UT2 telescope \citep{Vreeswijk2018}, with the value $z=0.654$. 
In the radio band, both the Arcminute Microkelvin Imager Large Array (AMI-LA) at 15.5 GHz \citep{GCN_radio1} and the Giant Metrewave radio Telescope at 1.4 GHz \citep{GCN_radio2} observed the field and detected the source.

\begin{figure} 
    \includegraphics[width=\columnwidth]{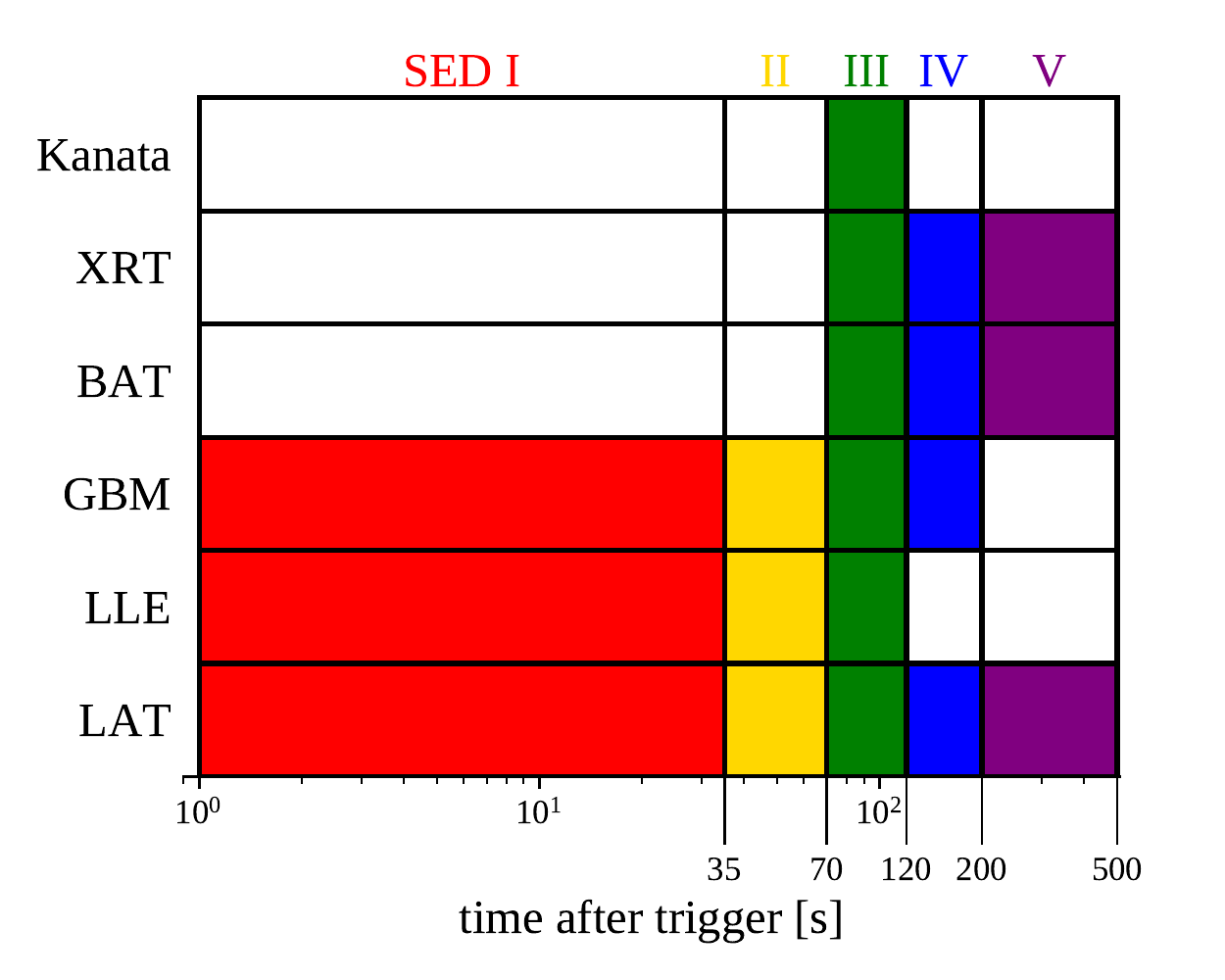}
    \caption{Time intervals and corresponding instruments providing the spectra used for the spectral analysis of GRB 180720B.  The color code corresponding to the sequence of time intervals is used for the following figures but  Fig.~\ref{fig:sed1} and Fig.~\ref{fig:sed1_cornerplot}.
   }
    \label{fig:timeline}
\end{figure}

In order to study the transition from the prompt to the afterglow emission in GRB~180720B, we considered five time intervals: 0--35\,s, 35--70\,s, 70--120\,s, 120--200\,s and 200--500\,s (see Fig. \ref{fig:timeline}). These time intervals have been identified with different colors (red, yellow, green, blue and purple, respectively) and this color-code has been used also in Fig.~\ref{fig:lc_GBM} and \ref{fig:lc_lat}. The choice of this subdivision is due both to the different starting times of the observations in different bands and to the variability of the light curves and spectral features of LAT high energy data. In particular, as Fig. \ref{fig:lc_GBM} shows, the first time interval (0-35\,s) includes the main emission event of the GRB. Furthermore, until 35 s the high energy LAT spectrum presents a different behaviour with respect to the rest of the burst (see lower panel in Fig.~\ref{fig:lc_lat}). This aspect will be discussed in \S \ref{sec:results} in more detail. The second time interval (35-70\,s) includes a secondary emission event visible in the GBM light curves (Fig. \ref{fig:lc_GBM}). The third one (70-120\,s) encompasses the peak of the LAT light curve (upper panel in Fig. \ref{fig:lc_lat}) and the last two intervals (120-200\,s and 200-500\,s) follow its decay.
In Fig.~\ref{fig:timeline}, for each time interval we mark the instruments we use for the analysis.
In the following sections we describe the main data source and the standard procedures adopted for the the data extraction. 
%-------------------------------------------------------------
%-------------------------------------------------------------

\subsection{\fe/LAT}

The LAT instrument on board \fe\ is composed by a tracker and a calorimeter sensitive to $\gamma$--ray photons in the energy range between $30$\,MeV and $300$\,GeV \citep{Ackermann2013}. 
Data extraction and analysis have been performed with \texttt{gtburst} which is distributed as part of the official \texttt{Fermitools} software\footnote{\url{https://fermi.gsfc.nasa.gov/ssc/data/}\label{Fermiwebsite}}. 
LAT data were extracted within a temporal window extending 900\,s after the trigger time and filtered selecting photons with energies in the 100\,MeV -- 100\,GeV range, within a region of interest (ROI) of $12^\circ$ centered on the burst position provided by \sw/BAT \citep{Barthelmy2018}. 
A further selection of photons with a zenith angle from the spacecraft axis $<100^\circ$ was applied in order to reduce the contamination of photons coming from the Earth limb. 
The highest photon energy is 4.9\,GeV and the corresponding photon has been detected by LAT 137 s after the GBM trigger. 
These values are consistent with what reported by the \fe/LAT Collaboration \citep{GCN_LAT}.

\subsection{\fe/GBM}\label{sec:fermi_gbm}

The GBM detector onboard \fe\ is composed of 12 sodium iodide (NaI, 8\,keV -- 1\,MeV) and two bismuth germanate (BGO, 200\,keV -- 40\,MeV) scintillation detectors \citep{Meegan2009}. 
We analysed the data from the two NaI detectors and the BGO detector with the highest count rates, namely NaI\#6, NaI\#8 and BGO\#1. 
For each detector, we retrieved the data and the most updated response matrix files from the \fe\ website\textsuperscript{\ref{Fermiwebsite}}. 

Spectra were extracted with \texttt{gtburst}: we selected energy channels in the range 8–900\,keV for NaI detectors, and 0.3–40\,MeV for the BGO detector, and excluded channels in the 30–40\,keV range due to the presence of the Iodine K-edge at 33.17\,keV\footnote{\url{https://fermi.gsfc.nasa.gov/ssc/data/analysis/GBM_caveats.html}}. 
The background was modelled over pre- and post-burst time intervals with a polynomial whose order is selected automatically by \texttt{gtburst}. 

We also fitted the first two time intervals by combining the GBM with the LAT Low Energy events (LLE) data \citep{Pelassa2010}. 
LAT-LLE data were retrieved from the \fe\ LLE Catalog\footnote{\url{https://heasarc.gsfc.nasa.gov/W3Browse/fermi/fermille.html}} and reduced with a similar procedure as for the GBM data through \texttt{gtburst}. 
The LLE spectra analysed cover the energy range 30-100\,MeV \citep[e.g.][]{Ajello2019}.

\subsection{\sw}
The Burst Alert Telescope (BAT) on-board \sw\ \citep{Gehrels2004} is a coded aperture mask which triggers GRBs by imaging photons in the energy range 15--350\,keV. 
We downloaded BAT event files from the \sw\ archive\footnote{\url{http://heasarc.gsfc.nasa.gov/cgi-bin/W3Browse/swift.pl}}. 
BAT spectra have been extracted with the {\tt batbinevt} task of {\sc heasoft} package (v6.25) and corrected for systematic errors (with the {\tt batupdatephakw} and {\tt batphasyserr} tasks). 
The response matrices have been computed by the {\tt batdrmgen} tool for the time intervals before, during and after the \sw\ slew. 

The X--Ray Telescope (XRT) focuses photons in the 0.3--10\,keV energy range \citep{Wells2004}. 
We downloaded the XRT event files from \sw/XRT archive\footnote{\url{http://www.swift.ac.uk/archive/}}. 
The source and background XRT spectra have been extracted with {\tt xselect} and standard procedures \citep{Romano_06} have been adopted to correct for the pile-up of X--ray photons.
We have generated the corresponding ancillary response files by the {\tt xrtmkarf} task. 
The energy channels of XRT below 0.5\,keV and above 10\,keV have been excluded. 
To include the XRT spectra to the joint broad-band spectral modeling, we have grouped the energy channels by the {\it grppha} tool to have at least 20 counts per bin. 

\subsection{Spectral analysis}\label{sec:spec_analysis}

Throughout the paper we refer to the GBM trigger time.
The LAT data have been extracted and analysed for all the five time intervals defined in Fig.~\ref{fig:timeline}. 
We used \texttt{gtburst} performing an unbinned likelihood analysis and assuming a power law model for the source photon spectrum\footnote{\url{https://fermi.gsfc.nasa.gov/ssc/data/analysis/scitools/gtburst.html}}. 
We included the \texttt{P8R3\_TRANSIENT020E\_2} instrument response function, a galactic model template and an isotropic template for particle background to take into account the background emission from the Milky Way, extra-galactic diffuse gamma-rays, unresolved extragalactic sources, residual (misclassified) cosmic-ray emission and other extragalactic sources\footnote{\url{https://fermi.gsfc.nasa.gov/ssc/data/analysis/documentation/Cicerone/Cicerone.pdf}}.

For the other instruments we used XSPEC (v12.10.0c) to perform a joint spectral analysis which combined any of the data sets of BAT, XRT, GBM or LLE in all the time intervals.  
In order to account for inter-calibration uncertainties between the different instruments, we introduced multiplicative factors in the fitting models. 
We left these factors free to vary except for the detector with the highest count rates, i.e. NaI\#6, whose factor has been frozen to 1. 
Inspecting the results of the fits the calibration factors between the GBM detectors agree within $\sim 15$\% and those between the GBM and the LLE amount to $\sim 35$\%.

The first three time intervals are fitted with a physical model for the synchrotron emission from relativistic electrons. 
We considered relativistic electrons injected with an energy distribution $dN(\gamma)/d\gamma \propto \gamma^{-p}$ between $\gamma_{\rm min}$ and $\gamma_{\rm max}$ and solved the continuity equation accounting for synchrotron losses. 
The spectral shape at any time $t$ is determined by the ratio between 
$\gamma_{\rm min}$ and the electron cooling energy $\gamma_{\rm c}$, and 
by the value of $p$. 
In addition to the spectral shape, the model spectrum is specified by two other parameters that determine respectively the energy $E_{\rm m}$ of the photons emitted by electrons of energy $\gamma_{\rm min}$, and the normalization $F_m$ of the $F(E)$ spectrum evaluated at $E_{\rm m}$.
Such a model is not present in the XSPEC library and this way to parametrize the model is appropriate for the implementation in XSPEC. Therefore we built table model spectra for different combination of the free parameters and implemented them in XSPEC. 

From the value of $E_{\rm m}$ and the ratio $\gamma_{\rm min}/\gamma_{\rm c}$ we can derive also the cooling energy $E_{\rm c}$ of the photons emitted by the electrons cooled down to $\gamma_{\rm c}$ using the relation $E_{\rm c} = E_{\rm m} \left(\gamma_{\rm c}/\gamma_{\rm min} \right)^2$. So the four free parameters can be redefined to be $E_{\rm c}$, $E_{\rm m}$, $p$ and the normalization $F_m$.

Confidence ranges on these parameters were then derived within XSPEC, through the built-in Markov Chain Monte Carlo algorithm (\texttt{chain} command). 
The results of the spectral analysis of all the data used in this work are reported in Table~\ref{tab:params}.

The four free parameters defined above are also associated to the five physical quantities describing the emitting region, namely the comoving magnetic field $B^\prime$, the total number of emitting electrons $N_{\rm e}$, the bulk Lorentz factor $\Gamma$ and the values of $\gamma_{\rm c}$ and $\gamma_{\rm min}$.
We derive the value of the bulk Lorentz factor $\Gamma$ from the onset of the afterglow (see \S \ref{sec:gamma0}).
The other four physical quantities can be constrained using this value of $\Gamma$ and the results of the spectral fitting (\S \ref{sec:discussion}).

%-------------------------------------------------------------
\begin{figure} 
%\hskip -0.5 cm 
% \centering
    \includegraphics[width=\columnwidth]{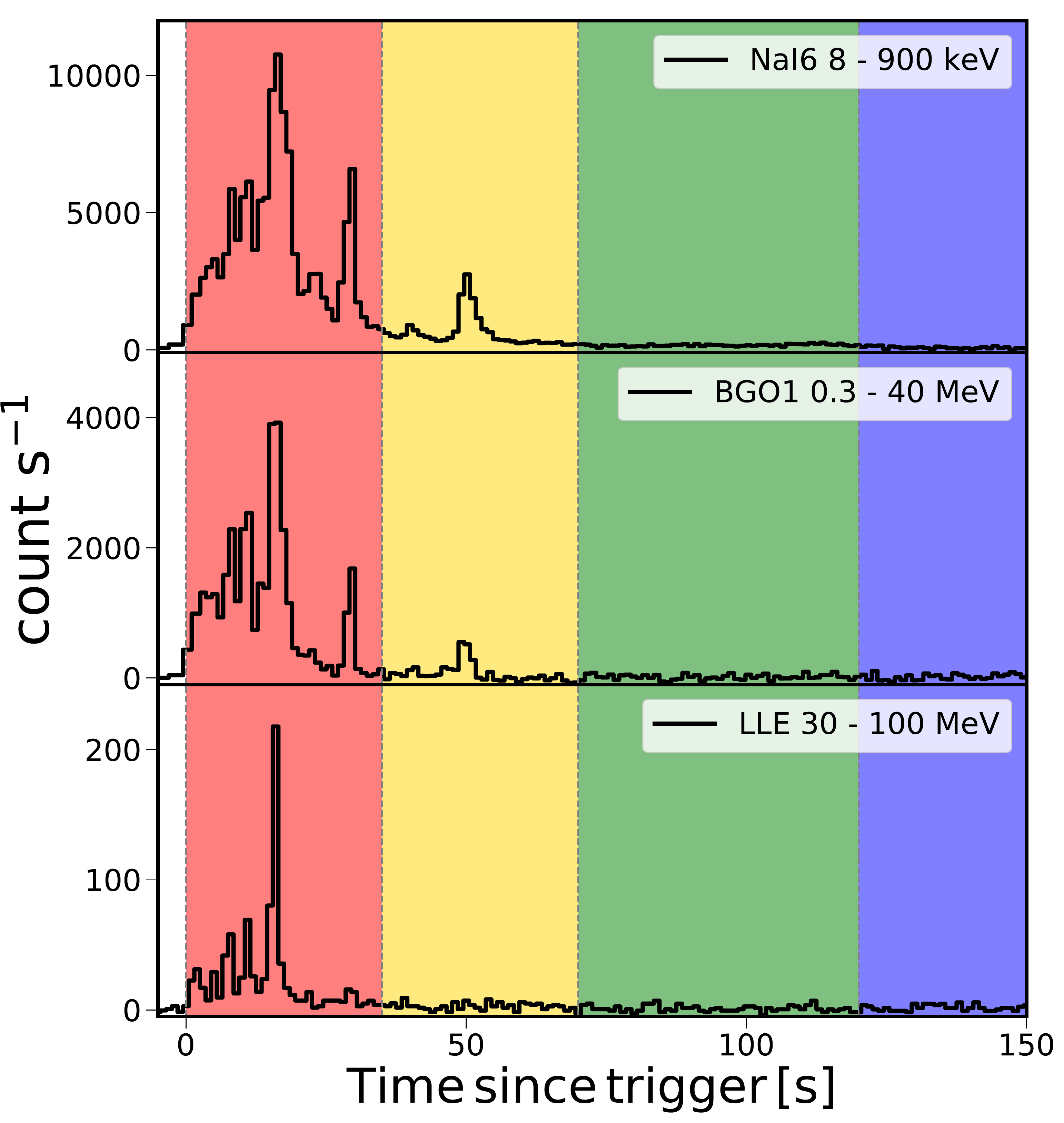}
    \caption{Background-subtracted light curves of GRB 180720B, detected by NaI\#6 (8--900\,keV, top), BGO\#1 (0.3--40\,MeV, middle), and LAT–LLE (30--100\,MeV, bottom). Different time intervals are highlighted with different colors, according to the color code used in Fig.~\ref{fig:timeline}. }
    \label{fig:lc_GBM}
\end{figure}
%-------------------------------------------------------------

%-------------------------------------------------------------
\begin{figure} 
%\hskip -0.5 cm 
% \centering
    \includegraphics[width=\columnwidth]{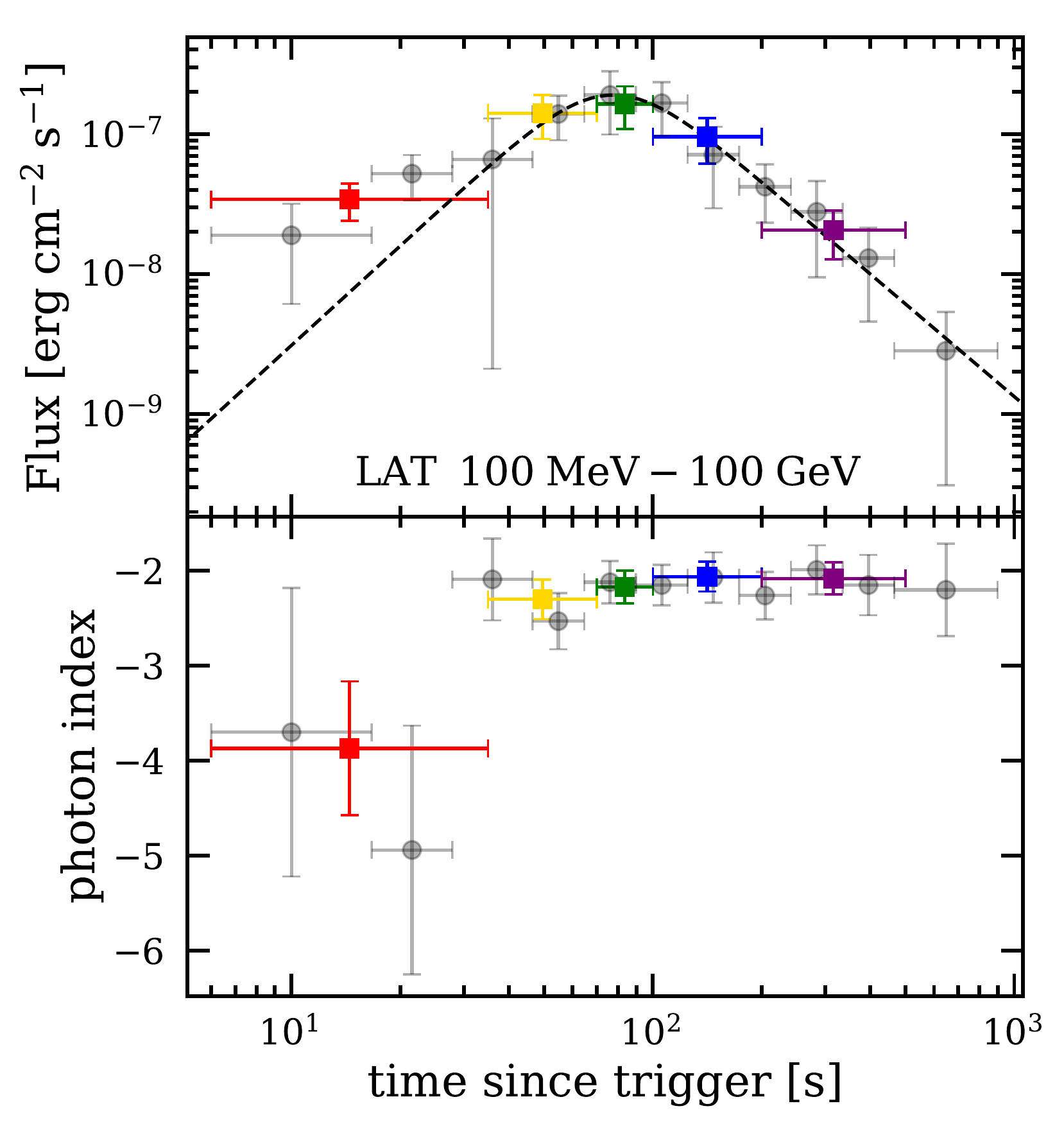}
    \caption{\textit{Top}: LAT light curve. The energy flux is integrated over the 100\,MeV -- 100\,GeV energy range. Colored square symbols correspond to the time intervals defined in \ref{fig:timeline}; gray circles show the results of the analysis with a higher time resolution. \textit{Bottom}: Photon indices of the power law model used for the LAT data analysis. Same color coding and symbols of the top panel.} 
    \label{fig:lc_lat}
\end{figure}
%-------------------------------------------------------------

\section{Results}\label{sec:results}

Fig.~\ref{fig:lc_GBM} shows the light curves (counts s$^{-1}$) of GRB 180720B detected by three different instruments sensitive to increasing photon energies from top to bottom, NaI (8--900\,keV), BGO (0.3--40\,MeV) and LLE (30--100\,MeV). 
These light curves show a main event lasting $\sim 35$ s with numerous overlapping pulses and a very bright peak at $t \simeq 15$ s, followed by another peak at $t \simeq 50$\,s. The pulses show the typical fast rise exponential decay (FRED) shape.
The observed burst duration $T_{90}$ is  49\,s in the energy range 50-300 keV \citep{Roberts2018}.   

Fig.~\ref{fig:lc_lat} shows the light curve (top panel) and the photon index evolution (lower panel) obtained from the analysis of the LAT data in the energy range 100 MeV - 100 GeV.
The colour-coded symbols correspond to the five time intervals defined in Fig.~\ref{fig:timeline} while the grey points show the results of the analysis of the LAT data on a finer temporal binning using equally-spaced logarithmic temporal bins, except for the first and last time bins, which are longer to have higher photon statistics. 
The delay between the first photon detected by LAT in the 100 MeV -- 100\,GeV  range and the GBM trigger is approximately 5\,s. 

To fit the LAT spectrum in each time interval we use a power law model. The results are resumed in Table~\ref{tab:params}. The lower panel of Fig.~\ref{fig:lc_lat} shows the evolution of the power law photon spectral index. 
The spectrum of the emission detected by LAT up to 35\,s (shown by the first two grey symbols or by the combined red square symbol) is characterized by a soft spectrum with spectral index $\Gamma_{\rm PL}< -3$. 
From 35\,s onward, the LAT photon index sets on the constant value $\sim -2$. 
This spectral change is confirmed also by a finer time-resolved analysis (grey points).
The softer spectrum $\Gamma_{\rm PL}\sim -4$ of the first time interval suggests that up to $\sim35$\,s the emission detected by LAT is the spectral tail of the prompt emission spectrum extending into the LAT energy range. 
On the other hand, the harder spectral slope $\sim$ -2 of the long--lived LAT emission after 35\,s is a common feature of LAT GRBs \citep{Ajello2019, Nava2018} and we interpret it as the emission produced by the external forward shock. 
As shown in Fig.~\ref{fig:lc_lat} the spectral index of $\sim$--2 remains constant until 900\,s, i.e. long after the prompt emission has ceased. This can be an indication that in LAT we are observing the transition from the prompt to the afterglow emission phases.

The LAT light curve (top panel of Fig.~\ref{fig:lc_lat}) is derived by integrating the spectrum over the energy range 100\,MeV -- 100\,GeV in each temporal bin. The light curve peaks at $t_{\rm peak}\sim78$\,s (observed frame): we interpret this as the onset time of the afterglow emission and infer the bulk Lorentz factor of the outflow in \S \ref{sec:gamma0}.

From the spectral analysis we can infer that the LAT light curve, in the 0--35\,s time interval, is possibly the superposition of a dominant prompt emission soft tail and a rising harder afterglow components. 
To characterize the time profile of only the afterglow component, we thus consider the light curve beyond 35\,s.
We fit it with a smoothly-joint double power law \citep[see e.g.][]{Ghirlanda2010}:
\begin{equation*}
    R(t) = \frac{A(t/t_\mathrm{b})^{\alpha}}{1+(t/t_\mathrm{b})^{\alpha+\beta}}
\end{equation*}
where the free parameters are the rise and decay slopes $\alpha$ and $\beta$ respectively, the characteristic time $t_\mathrm{b}$ and the normalization factor $A$. 
The best--fit parameters are $A=(3.8 \pm 0.4) \times10^{-7}$\,erg\,s$^{-1}$\,cm$^{-2}$, $\alpha=2.4\pm 0.7$, $\beta=2.2\pm0.2$ and $t_\mathrm{b}=76.5\pm8.0$\,s. 
The peak of the light curve is given by $t_\mathrm{peak} = t_\mathrm{b} (\alpha/\beta)^{1/(\alpha+\beta)} \sim 78$\,s. 
This fit is shown by the dashed line in the top panel of Fig.~\ref{fig:lc_lat}. 

The LAT flux rises consistently with $\sim t^2$, that is the expected behaviour in case of synchrotron emission (in the fast cooling regime) from the external shock prior to the deceleration radius in a constant ambient medium \citep[e.g.][]{Sari1999}. 
The flux decay follows a $\sim t^{-2}$ trend which is somewhat 
unusual\footnote{\cite{Ackermann2013} and \cite{Ajello2019} reported an average temporal decay index around $-1$ for LAT GRB light curves.},
but consistent with what derived by  \cite{Ajello2019} ($-1.88 \pm 0.15$).

After 78 s, the LAT flux is a significant fraction of the total emission and characterized by a nearly flat spectrum in $E F(E)$. 
This implies that at this time the LAT flux can be considered a proxy of the bolometric flux. \cite{Ghisellini2010} and \cite{Ackermann2013} studied the theoretical temporal evolution of the observed bolometric flux emitted by a fireball expanding in a homogeneous interstellar medium. They found that the flux decays as $\sim t^{-1}$ in the adiabatic regime and as $\sim t^{-10/7}$ in the radiative regime.
The decay slope we find is steeper than both these theoretical predictions. 

This unusual behavior has been investigated by \cite{Panaitescu2017}, who found that the LAT light curves of bright bursts (with $>10^{-4}$\,ph\,s$^{-1}$) tend to show a steeper temporal decay at early stages.

\subsection{Spectral evolution}

\begin{figure}
%\hskip -0.4 cm 
% \centering
    \includegraphics[scale=0.47]{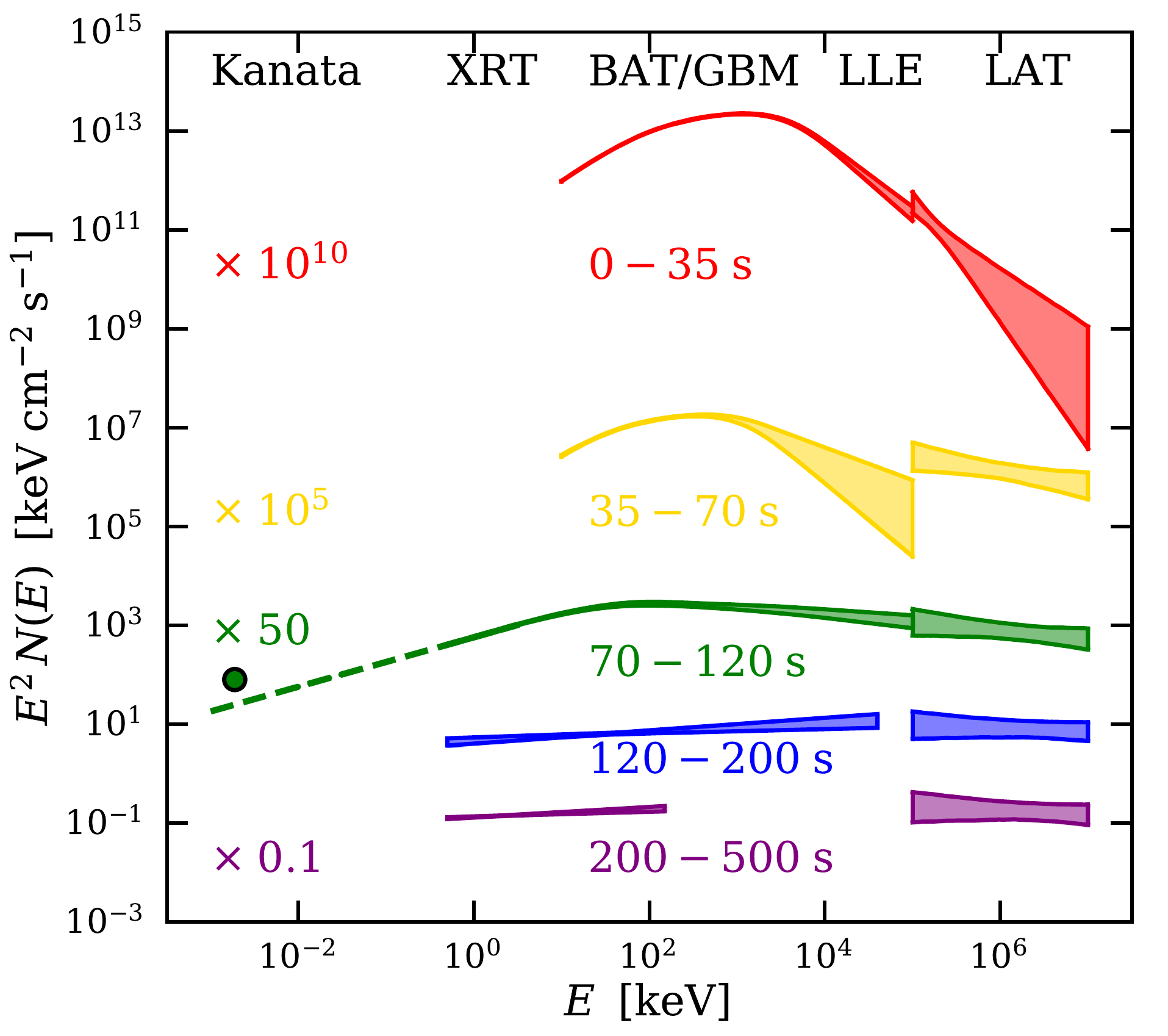}
    \caption{Evolution of the spectral energy distribution of GRB~180720B. Each curve corresponds to a specific time interval and it has been rescaled for presentation purposes by the scaling factor reported in the figure. For reference, the top labels mark the instruments providing data in the corresponding energy ranges.}
    \label{fig:sed}
\end{figure}

Fig.~\ref{fig:sed} shows the evolution of the SED up to 500 seconds. This plot shows the SED obtained from the combined analysis of the XRT, GBM and LAT/LLE data when available (first three SED) together with the independent analysis of the LAT data (for all the five time intervals). Since the photon with the highest energy detected by LAT is $\sim5$\,GeV, we represent the LAT SEDs limited over the 0.1--10\,GeV energy range.

The shaded regions corresponding to each SED show the 68\% confidence interval on the model fits. These are obtained by combining the errors on the best fit parameters and accounting for their covariance through a Monte Carlo sampling of the parameter space. In the following sections, we describe the evolution of the SEDs of GRB 180720B.

%-------------------------------------------------------------
%-------------------------------------------------------------

\subsubsection{SED I (0-35 sec): evidence of synchrotron prompt spectrum}

The first SED, corresponding to the time interval 0--35 seconds, is represented in red in Fig.~\ref{fig:sed} and shows a spectrum which extends from 10\,keV all through the LAT energy range with a single emission component. The independent analysis of the LAT data (as discussed above) with a power law results in a soft emission component with slope $\Gamma_{\rm PL}=-3.87 \pm 0.71$. A similar spectral slope value is obtained from the analysis of the LLE data in the 30--100\,MeV range. These spectral results are reported in Table \ref{tab:params}.

\begin{figure*}
    \centering
    \includegraphics[width=\textwidth]{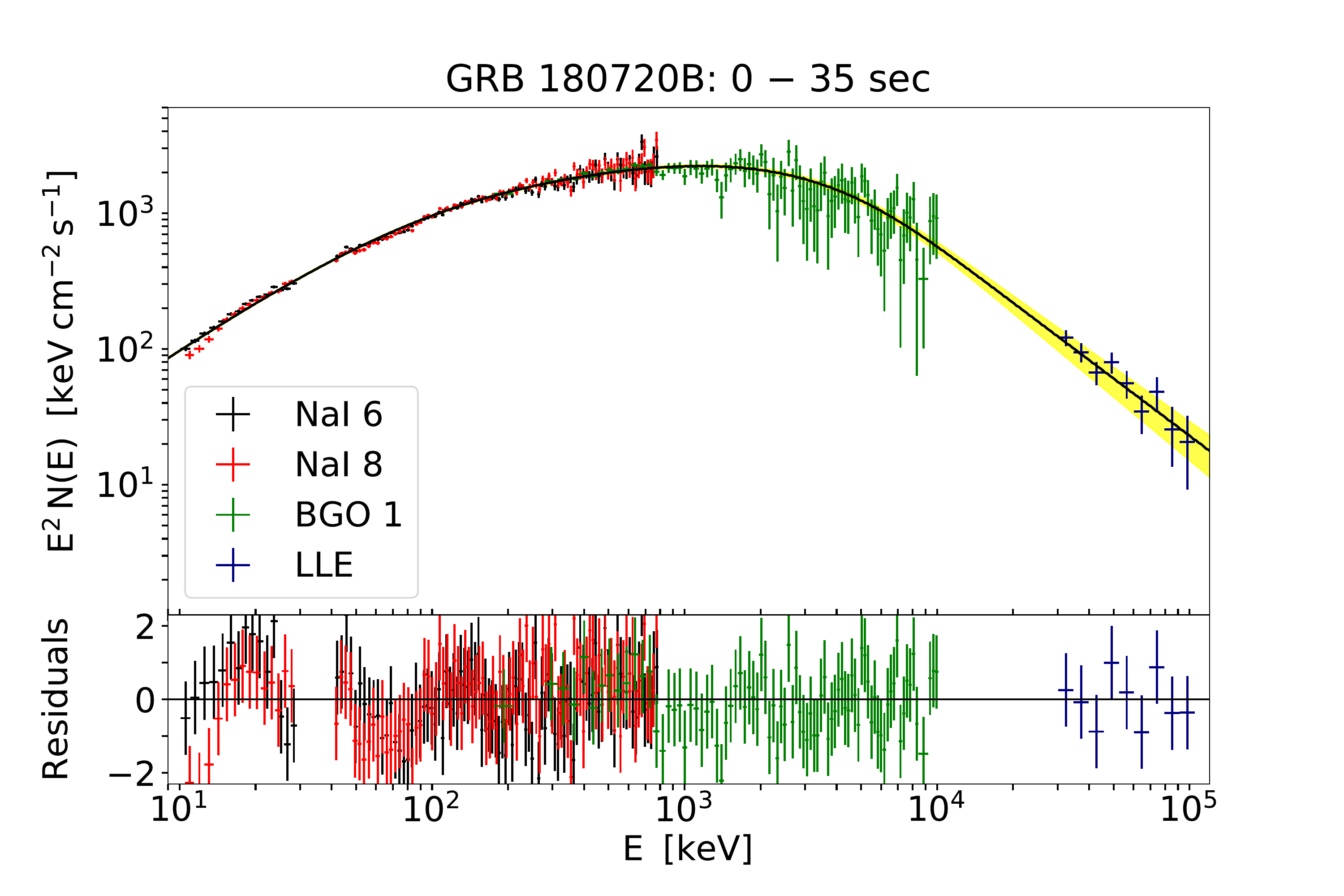}
    \caption{Spectrum corresponding to the 0-35 s time interval. The GBM and LLE data are fitted jointly with the model (shown by the solid line) of synchrotron emission from relativistic electrons. The shaded yellow region represents the 68\% confidence region of the best fit model. The bottom panel shows the data--to--model residuals. The best fit spectral parameters are reported in Table\ref{tab:params}.}
    \label{fig:sed1}
\end{figure*}

\begin{figure}
    \centering
    \includegraphics[width=\columnwidth]{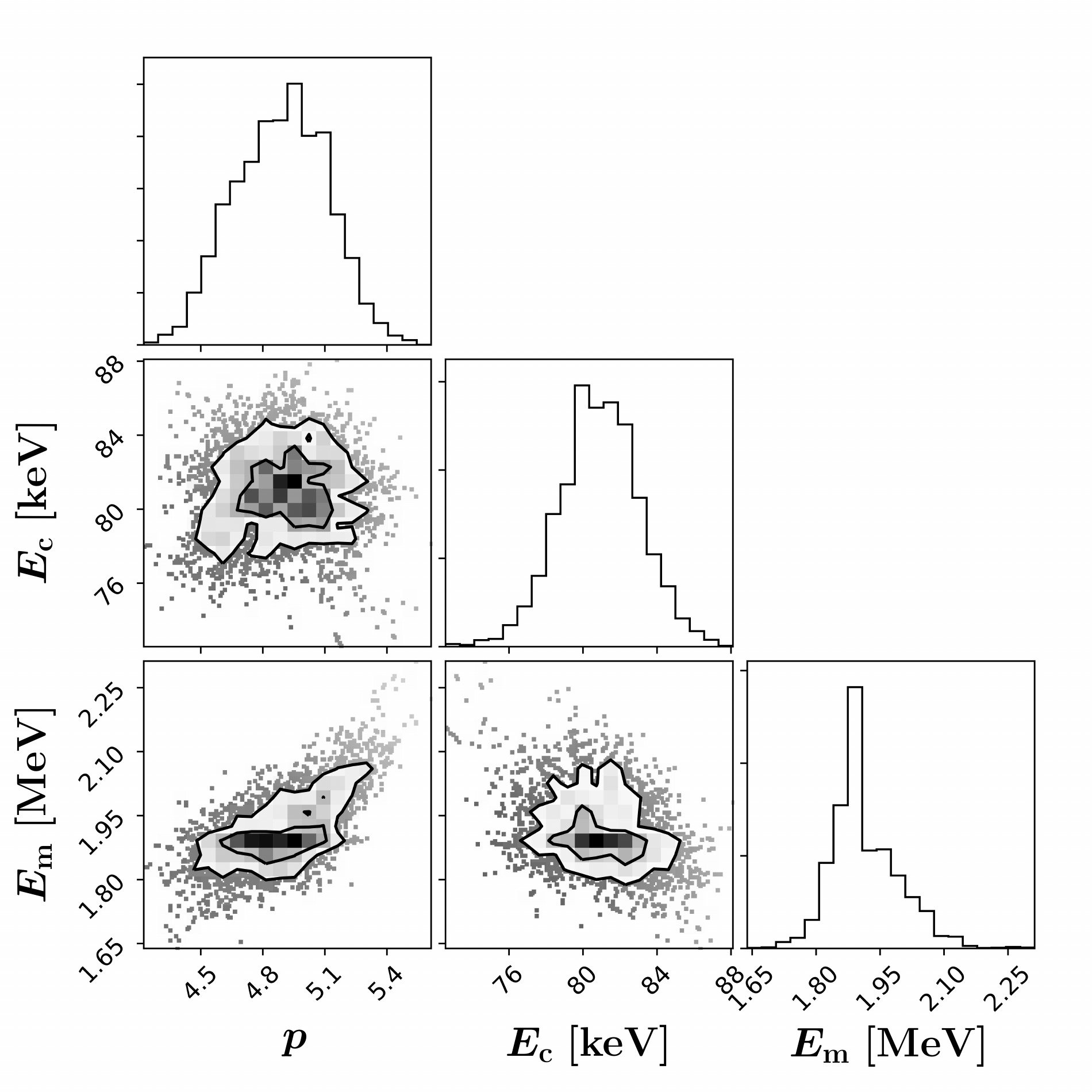}
    \caption{Corner plot of the free parameters of the synchrotron model used to fit the SED I. The posterior distributions and their correlations in the two dimensional plots are shown. }
    \label{fig:sed1_cornerplot}
\end{figure}

We fitted the GBM (NaI+BGO) and LLE data together with the synchrotron model we implemented in XSPEC. The spectral parameters are reported in Table~\ref{tab:params}. 
The best fit model is shown in Fig.~\ref{fig:sed1} by the solid black line and the shaded yellow region represents the 68\% confidence interval. The corner plot of the posterior distributions of the parameters of the fit is shown in Fig.~\ref{fig:sed1_cornerplot}, where no strong residual correlation between the free parameters is evident. The residuals of the fit (bottom panel of Fig.~\ref{fig:sed1}) show that the model properly fits the data over almost the entire spectral range. 
Systematic residuals are present below $\sim$30\,keV over a narrow energy range. 
The nature of such residuals could be due to poorly calibrated response and/or to the break evolution within the considered time bin \citep{Ravasio2019}.

The energy spectrum ($F(E) = E N(E)$) shows a slope $F(E) \propto E^{1/3}$ before the energy break $E_{\rm b} \sim E_{\rm c}$ and a slope $F(E)\propto E^{-1/2}$ between the break and the peak energy $E_{\rm p} \sim E_{\rm m}$.
These results, similar to what recently found in other GRBs detected by \sw\ and \fe\ with either empirical (e.g. \citealt{Zheng2012, Oganesyan2017, Oganesyan2018, Ravasio2018, Ravasio2019}) or with physically motivated synchrotron models \citep{Zhang_2016, Burgess2018, Oganesyan2019}, suggest that the emission is in the fast cooling regime \citep{Kumar2008, Daigne2011, Beniamini2013}, even if the cooling is not complete.

The best fit model returns a steep slope $p\sim 4.8$ of the injected electron distribution. 
In empirical models (e.g. Band model) this corresponds to a spectral photon index $\beta =-p/2 -1 \sim -3.4$ which is consistent with the value obtained from the independent fit of the LLE data alone. 
The implication of the relatively steep $p$ value found will be discussed in \S \ref{sec:discussion}.

%-------------------------------------------------------------
%-------------------------------------------------------------
%-------------------------------------------------------------
%-------------------------------------------------------------

%---------------------------------------------------------------
\begin{table*}
\centering 
\caption{Best-fit parameters for each different model used in the five SEDs reported in Fig.~\ref{fig:sed} of GRB 180720B. The first column reports the time interval over which the spectrum has been integrated, the second column the detector whose data have been used to build the SED along with the energy range and the third column represents the different models used to fit that data. From the fourth column onwards: energy flux computed in the energy range reported in the second column, photon energy corresponding to the electron cooling Lorentz factor $E_\mathrm{c}$, photon energy corresponding to the electron injection Lorentz factor $E_\mathrm{m}$, electron energy distribution slope $p$ (corresponding high-energy photon index $\beta = -p/2 -1$ in parentheses), power law photon index of the LLE/LAT data, fit p-value where available. }

\label{tab:params}

\begin{adjustbox}{max width=\textwidth}
\begin{tabular}{ccccccccccc}
\hline\hline
  \multicolumn{1}{c}{\multirow{2}{*}{Time interval}} &
  \multicolumn{1}{c}{\multirow{2}{*}{Data}} &
  \multicolumn{1}{c}{\multirow{2}{*}{Model}} &
  \multicolumn{1}{c}{Flux} &
  \multicolumn{1}{c}{$E_{\rm c}$} &
  \multicolumn{1}{c}{$E_{\rm m}$} &
  \multicolumn{1}{c}{\multirow{2}{*}{$p(\beta)$}} &
  \multicolumn{1}{c}{\multirow{2}{*}{$\Gamma_{\rm PL}$}} &
  \multicolumn{1}{c}{\multirow{2}{*}{Prob}}  \\
 & & & [$\rm 10^{-7} erg / s\, cm^{2}$] & [keV] & [keV] & & & & & \\
\hline
  \multirow{8}{*}{0-35 s} & GBM & \multirow{2}{*}{Sync} & \multirow{2}{*}{$136_{-1.}^{+3.}$} & \multirow{2}{*}{$81_{-10}^{+13}$} & \multirow{2}{*}{$1894_{-230}^{+54}$} & \multirow{2}{*}{$4.79_{-0.23}^{+0.81} (-3.39_{-0.40}^{+0.12})$} & \multirow{2}{*}{-} & \multirow{2}{*}{0.92}\\
         & [10 keV - 40 MeV] & & & & & & & \vspace{0.1cm} \\
         & LLE & \multirow{2}{*}{PL} & \multirow{2}{*}{$1.73_{-0.26}^{+0.16}$} & \multirow{2}{*}{-} & \multirow{2}{*}{-} & \multirow{2}{*}{-} & \multirow{2}{*}{$-3.44_{-0.21}^{+0.18}$} & \multirow{2}{*}{0.96} \\
         & [30 - 100 MeV] & & & & & & & \vspace{0.1cm} \\
         & GBM+LLE & \multirow{2}{*}{Sync} & \multirow{2}{*}{$143_{-3}^{+2}$} & \multirow{2}{*}{$81_{-6}^{+6}$} & \multirow{2}{*}{$1883_{-106}^{+39}$} & \multirow{2}{*}{$4.76_{-0.11}^{+0.36}$ $(-3.38_{-0.18}^{+0.06})$} & \multirow{2}{*}{-} & \multirow{2}{*}{0.96} \\
         & [10 keV - 100 MeV] & & & & & & & \vspace{0.1cm} \\
         & LAT & \multirow{2}{*}{PL} & \multirow{2}{*}{$0.34_{-0.10}^{+0.10}$} & \multirow{2}{*}{-} & \multirow{2}{*}{-} & \multirow{2}{*}{-} & \multirow{2}{*}{$-3.87_{-0.71}^{+0.71}$} & \multirow{2}{*}{-} \\
         & [100 MeV - 100 GeV] & & & & & & & \\
         \hline
  \multirow{4}{*}{35-70 s} & GBM+LLE & \multirow{2}{*}{Sync} & \multirow{2}{*}{$13.0_{-2.4}^{+0.14}$} & \multirow{2}{*}{$25_{-7}^{+3}$} & \multirow{2}{*}{$520_{-66}^{+30}$} & \multirow{2}{*}{$3.38_{-0.06}^{+1.65}$  $(-2.69_{-0.83}^{+0.03})$} & \multirow{2}{*}{-} & \multirow{2}{*}{0.82} \\
        & [10 keV - 100 MeV] & & & & & & & \vspace{0.1cm} \\
        & LAT & \multirow{2}{*}{PL} & \multirow{2}{*}{$1.41_{-0.49}^{+0.49}$} & \multirow{2}{*}{-} & \multirow{2}{*}{-} & \multirow{2}{*}{-} & \multirow{2}{*}{$-2.3_{-0.21}^{+0.21}$} & \multirow{2}{*}{-} \\
        & [100 MeV - 100 GeV] & & & & & & & \\
        \hline
  
   \multirow{4}{*}{70-120 s} & XRT+BAT+GBM+LLE & \multirow{2}{*}{Sync} & \multirow{2}{*}{$7.26_{-0.8}^{+0.21}$} & \multirow{2}{*}{-} & \multirow{2}{*}{$43.55_{-7.9}^{+10.48}$} & \multirow{2}{*}{$2.08_{-0.04}^{+0.19}$ $(-2.04_{-0.02}^{+0.1})$} & \multirow{2}{*}{-} & \multirow{2}{*}{0.92} \\
           &[0.5 keV - 100 MeV] & & & & & & & \vspace{0.1cm} \\
           & LAT & \multirow{2}{*}{PL} & \multirow{2}{*}{$1.64_{-0.56}^{+0.56}$} & \multirow{2}{*}{-} & \multirow{2}{*}{-} & \multirow{2}{*}{-} & \multirow{2}{*}{$-2.17_{-0.17}^{+0.17}$} & \multirow{2}{*}{-} \\
           & [100 MeV - 100 GeV] & & & & & & & \\
        \hline
  \multirow{4}{*}{120-200 s} & XRT+BAT+GBM & \multirow{2}{*}{PL} & \multirow{2}{*}{$1.38_{-0.15}^{+0.20}$} & \multirow{2}{*}{-} & \multirow{2}{*}{-} & \multirow{2}{*}{-} & \multirow{2}{*}{$-1.91 \pm 0.05$} & \multirow{2}{*}{0.98} \\
           & [0.5 keV - 40 MeV] & & & & & & & \vspace{0.1cm} \\
           & LAT & \multirow{2}{*}{PL} & \multirow{2}{*}{$0.96_{-0.34}^{+0.34}$} & \multirow{2}{*}{-} & \multirow{2}{*}{-} & \multirow{2}{*}{-} & \multirow{2}{*}{$-2.06_{-0.16}^{+0.16}$} & \multirow{2}{*}{-} \\
           & [100 MeV - 100 GeV] & & & & & & & \\
        \hline
  \multirow{4}{*}{200-500 s} & XRT+BAT & \multirow{2}{*}{PL} & \multirow{2}{*}{$0.06 \pm 0.01$} & \multirow{2}{*}{-} & \multirow{2}{*}{-} & \multirow{2}{*}{-} & \multirow{2}{*}{$-1.86 \pm 0.03$} & \multirow{2}{*}{0.33} \\
           & [0.5 - 150 keV] & & & & & & & \vspace{0.1cm} \\
           & LAT & \multirow{2}{*}{PL} & \multirow{2}{*}{$0.21_{-0.08}^{+0.08}$}  & \multirow{2}{*}{-} & \multirow{2}{*}{-} & \multirow{2}{*}{-} & \multirow{2}{*}{$-2.08_{-0.17}^{+0.17}$} & \multirow{2}{*}{-} \\
           & [100 MeV - 100 GeV] & & & & & & & \\
        \hline
\hline\end{tabular}
\end{adjustbox}
\end{table*}
%---------------------------------------------------------------

\subsubsection{SED II (35-70 sec): transition between prompt and afterglow emission}
The second SED, represented in yellow in Fig.~\ref{fig:sed}, shows the presence of two emission components. The spectrum below 100\,MeV (obtained combining GBM and LLE data up to 100\,MeV) is still fitted by the synchrotron model albeit with a harder slope of the electron energy distribution $p \sim 3.4$. The values of the peak and break energies are a factor $\sim$ 2 smaller than those of SED I. The LAT data show a harder spectrum than in the previous SED ($\Gamma_{\rm PL} = -2.3 \pm 0.2$). Note that in the following time intervals the photon index of the LAT spectrum is even harder, settling around $\Gamma_{\rm PL} \sim -2$ (as reported in Fig.~\ref{fig:lc_lat}).
This suggests that we are observing the rise of the afterglow emission component at very high energies which is characterized by a typical photon index $\Gamma_{\rm PL} \sim -2$. The superposition of this harder emission component with the peaked prompt emission spectrum as seen by the GBM can account for the harder $p$ value obtained from the fit of the GBM+LLE data with the synchrotron model. 

In conclusion, the SED in the time interval 35-70 s shows the coexistence of two different components:
the prompt synchrotron emission dominates in the GBM+LLE energy range; above 100 MeV the LAT spectrum flux is inconsistent with the low energy spectrum flux and this could indicate that the afterglow component is rising and contributing at higher energies.

\subsubsection{SED III, IV, V: evidence of afterglow emission}

The time interval of the green SED from 70 s to 120 s contains the peak of the LAT light curve ($t_{\rm peak} \sim 78$ s, Fig.~\ref{fig:lc_lat}) and represents the moment when the afterglow begins to dominate the observed emission. This fact is also supported by the weak signal in the BAT and GBM energy range and by the absence of bumps in the GBM lightcurves (see Fig. \ref{fig:lc_GBM}). At this epoch, X--ray data from \sw/XRT ranging from 0.5 keV to 10 keV are also available, allowing to extend the analysis to lower energies. 
We use the synchrotron model to fit the combined data of XRT+BAT+GBM+LLE. The best fit is a synchrotron spectrum in fast cooling, with  $E_c$ constrained to be below 0.5 keV, $E_{\rm m} \sim 44$\,keV and a high energy slope which is consistent with the LAT spectrum (see also Table~\ref{tab:params}). The spectrum shows no evidence of a break down to the XRT energy range. Therefore, under the assumption that the emission is still in the fast cooling regime, at this epoch we expect that the cooling energy $E_{\rm c}$ is very low, less than 0.5 keV. Furthermore, if we extrapolate the spectrum down to the optical range with a power law $N(E) \propto E^{-3/2}$ (dashed green line in Fig.~\ref{fig:sed}), it is marginally consistent with the optical detection reported by \citealt{GCNottico}.

In the remaining two SEDs from 120\,s up to 500\,s the afterglow emission is dominant. Indeed data from X-rays up to GeV are well fitted by a single power law function with $\Gamma_{\rm PL} \sim -2$, which is the expected value of the synchrotron afterglow spectral photon index \citep{Burrows2005, Zhang2006}. 

\section{Estimate of the bulk Lorentz factor $\Gamma_0$}\label{sec:gamma0}

The light curve of the LAT flux  (Fig.~\ref{fig:lc_lat} top panel) shows a peak at $\sim$ 80\,s from the trigger. After 35-40 s,  the prompt 
emission is already too weak to contribute substantially to the observed emission in the LAT energy range. Indeed, the spectral slope of the LAT emission changes from soft to hard (Fig.~\ref{fig:lc_lat}  bottom panel). 

If the long--lasting LAT emission is synchrotron radiation from the external shock, we can derive the bulk Lorentz factor just prior the deceleration phase, conventionally $\Gamma_0$, from the interpretation of the peak of the LAT light curves as the deceleration time. The different derivation of $\Gamma_0$ proposed in the literature have been summarised and compared recently in \citet{Ghirlanda2018} where it has been shown that the different methods differ at most by a factor of 2. We therefore choose the equation derived by  \citet{Nava2013} :
\begin{equation} 
\Gamma_{0} 
= \left[\frac{(17-4s)(9-2s)3^{2-s}}{2^{10-2s}\pi(4-s)}
\left(\frac{E_0}{n_0 m_{p}c^{5-s}}\right)\right]^{1/(8-2s)}
t_{{\rm p}, z}^{-\frac{3-s}{8-2s}}
\label{n13}
\end{equation}
Here $t_{{\rm p},z}$ is the rest frame onset time, i.e. $t_{{\rm p},z} = t_{\rm p}/(1+z)$ and $m_{p}$ is the proton mass. 
The kinetic energy of the fireball is inferred as  $E_0 = E_{\rm iso}(1-\eta)/\eta$, i.e. the left--over of the prompt emission ($\eta$ is the efficiency of conversion of the initial energy into radiation during the prompt phase, typically assumed to be of few tens percent). 
The radial density profile is parametrized as $n = n_0 R^{-s}$, where $R$ is the distance from the central engine.
We  consider the uniform density ($s=0$) case and the scenario of a stellar wind density profile ($s=2$).
In the wind case $n_0=\dot M_{\rm w}/(4\pi v_{\rm w} m_{p})$, where $\dot M_{\rm w}$ is the rate of mass loss and $v_{\rm w}$ is the wind speed.

Assuming a redshift $z=0.654$ as reported in \cite{Vreeswijk2018}, an isotropic equivalent energy $E_{\rm iso} = 6\times 10^{53}$\,erg \citep{Frederiks2018}, $\eta=0.2$, $t_{\rm p}=80$\,s, we estimate $\Gamma_0 = 312~(234)$ for $n_0=1~(10)$\,cm$^{-3}$ in the case of a constant external medium density ($s=0$ in Eq.~\ref{n13}). For a wind medium ($s=2$ in Eq.~\ref{n13}), assuming a wind mass--loss rate $\dot M_{\rm w}=10^{-5} \; \rm{M_\odot \; yr^{-1}}$, we obtain $\Gamma_0 = 153~(86)$ for $v_{\rm w}=10^2~(10^{3})$\,km s$^{-1}$. Such values are consistent with the distributions of $\Gamma_0$ in both scenarios obtained from the analysis of a large sample of bursts with measured onset time \citep{Ghirlanda2018}. 

\section{Discussion}\label{sec:discussion}

\subsection{Prompt emission}

The synchrotron model presented in this work provides an acceptable fit of the prompt emission spectra of GRB~180720B in the energy range between 10\,keV and 100\,MeV. 
Under the assumption of one-shot electron acceleration, we can derive the physical parameters of the emission region, i.e. the magnetic field $B'$, the minimum energy of the injected non--thermal distribution of relativistic electrons $\gamma_{\rm min}$ and the total number of electrons $N_e$ contributing to the observed emission, from the spectral properties obtained from the fit, in particular the cooling energy $E_{\rm c}$, the injection energy $E_{\rm m}$, and the flux density at the cooling energy $F_{\rm c}$. 

Following \cite{Kumar2008} (see also \citealt{Beniamini2013}) we use the set of equations from \cite{Oganesyan2019}: 

\begin{equation}
    E_{\rm c} =  \frac{27 \pi \, e \, h \, m_e \, c \, (1+z)}{\sigma_T^2 \, B'^3 \, t_{\rm c}^2 \, \Gamma } 
\end{equation}

\begin{equation}
    E_{\rm m} =  \frac{3 \, e \, h \, B' \, \gamma_{\rm min}^2 \, \Gamma }{4 \pi \, m_e \, c  \, (1+z)}
\end{equation}

\begin{equation}
    F_{\rm c} = \frac{\sqrt{3} \, e^3 \, B' \, N_e \, \Gamma \, (1+z) }{4 \pi d_L^2 \, m_e \, c^2 }
\end{equation}
where $d_L$ is the luminosity distance of the GRB, $t_{\rm c}$ is the cooling time of electrons losing their energy via synchrotron radiation (neglecting synchrotron self Compton for simplicity) and $\Gamma$ is the bulk Lorentz factor. $F_{\rm c}$ is the flux at $E_{\rm c}$. Therefore, we can find the unknowns  $B'$, $\gamma_{\rm min}$ and $N_e$ in terms of the observables: 

\begin{align}
     B' &= \left( \frac{27 \pi \, e \, h \, m_e \, c \, (1+z)}{\sigma_T^2 \, E_{\rm c} \, t_{\rm c}^2 \, \Gamma } \right)^{1/3} \nonumber \\
     &\simeq 10 \; E_{\rm c,2}^{-1/3} \, t_{\rm c}^{-2/3} \, \Gamma_{2}^{-1/3} \, (1+z)^{1/3} \; \rm G
\end{align}
   
\begin{align}
    \gamma_{\rm min} &= \left( \frac{4 \pi \, m_e \, c \, E_{\rm m} \, (1+z)}{3 \, h \, e \, B' \, \Gamma } \right)^{1/2}  \nonumber \\
    &\simeq 6.3 \times 10^5 \; E_{\rm m,3}^{1/2} \, B'^{-1/2} \, \Gamma_{2}^{-1/2} \, (1+z)^{1/2}
\end{align}

\begin{align}
    N_e &= \frac{4 \pi d_L^2 \, m_e \, c^2 \, F_{\rm c}}{\sqrt{3} \, e^3 \, B' \, \Gamma \, (1+z) } \nonumber \\
    &\simeq 10^{50} \; F_{\rm c,\rm mJy} \, B'^{-1} \, \Gamma_{2}^{-1} \, d_{L,28}^2 \, (1+z)^{-1}
\end{align}
where $E_{\rm c}$ has been expressed in units of $10^2$\,keV, $E_{\rm m}$ in units of $10^3$\,keV, $\Gamma$ in units of 100, $F_{\rm c}$ in mJy and $d_L$ in units of $10^{28}$ cm.  

The fit of the spectrum corresponding to SED I (see Fig.~\ref{fig:sed1}) returns a cooling energy  $E_{\rm c} \sim 81$\,keV and an injection energy $E_{\rm m} \sim 1880$\,keV. 
We  derive also the flux at the cooling energy $F_{\rm c} \sim 6$ mJy. 
The integration time of 35\,s of SED I corresponds to the time (in the observer frame) needed by the injected electrons to cool down to $\gamma_{\rm c}$. The emission of these electrons produces a spectral break at an observed energy similar to the cooling energy $\sim E_{\rm c}$.
Therefore, we use  $E_{\rm c}=81$\,keV and $t_{\rm c}=35$\,s in Eq.~5. We assume the value of $\Gamma=\Gamma_0 \sim 300$ as estimated in the previous section for a homogeneous medium. 
Recalling that the redshift of GRB~180720B is $z = 0.654$, we find $B' \sim 1$ G, $\gamma_{\rm min} \sim 10^6$ and $N_e \sim 10^{51}$. These values are consistent with the ranges estimated by \cite{Oganesyan2019} for a sample of 21 \sw GRBs.

The best fit value from SED I for the electron spectral index is $p=4.8$
(Fig.~\ref{fig:sed1}). This in turn produces a soft photon spectrum  ($\beta  \sim -3.4$) at high energies as  confirmed by the independent fit of the high energy spectrum with a single powerlaw. 

Mildly relativistic shocks, produced e.g. by the dissipation of the kinetic energy of different colliding shells (internal shock scenario \citealt{Rees1994}), can hardly produce large values of $p$ and $\gamma_{\rm min}$. 
An efficient shock, as expected in low magnetized\footnote{ the magnetization $\sigma$ is defined as the ratio between the energy densities in magnetic field and in particles respectively, evaluated in the comoving frame of the emission region.} plasma \citep[e.g.][]{Sironi2011, Sironi2015b},  can accelerate particles to large $\gamma_{\rm min}$,  but typically converts a fraction (around 10\%) of the kinetic energy into a hard ($p\sim 2-2.4$)  non thermal electron energy distribution \citealt{Heavens1988, Rees1994, Kennel1984}. 

Magnetic reconnection, \citep{Spruit2001, Drenkhahn2002} as also recently shown by Particle In Cell simulations \citep{Sironi2014, Sironi2015a, Petropoulou2019}, can produce  
a steep electron energy distributions (i.e. $p\sim 4-5$) if the burst outflow has a high pair-proton number ratio ($\kappa \sim 200$) and a moderate magnetization parameter $\sigma\sim 1$. However, with such parameter values, the 
electrons can only attain a moderately large $\gamma_e \sim 10^2-10^3$. Different combinations of the leading parameters, though, could reconcile these values with those estimated in our analysis. 

If we compare the first SED I (0-35 s) with the second SED II (35-70 s), we find that the value of $p$ decreases from $p \sim 4.8$ to $p \sim 3.4$, thus corresponding to a harder electron energy distribution producing the synchrotron spectrum of SED II. 
However, we cannot exclude that also in SED II the value of $p$ is uncommonly large, due to the contamination of the high energy part of the spectrum by the harder spectral component arising in the LAT energy range. 

In conclusion, the acceleration mechanism responsible for the injection of non-thermal electrons must be really efficient in accelerating electrons to very high energies, but also it should give rise to a steep electron energy distribution.

We note that the large values of $\gamma_{\rm m}$ and  $\gamma_{\rm c}$ ensure that the self Compton emission occurs in the Klein--Nishina regime thus limiting the IC component \citep{Oganesyan2019}.

\subsection{LAT constraints on emission at 10 h}

The H.E.S.S. telescope detected high-energy photons ($\sim 300$\,GeV) from GRB~180720B at $\sim 10.5$\,h after the trigger time. \cite{Wang2019} report a H.E.S.S. flux of  $5\times 10^{-11}$\,erg\,cm$^{-2}$\,s$^{-1}$ in the energy range 100-440\,GeV. This very high energy emission can be the Synchrotron Self-Compton (SSC) component \citep{Meszaros1994,Waxman1997,Wei1998}. However, no detection by \fe/LAT was reported at this epoch.

By analysing the LAT data from 8 to 12 hours after the trigger, we obtain\footnote{We assume that the spectral index of the LAT emission component is -2 as observed before.} a 1 $\sigma$ upper limit of $8.5\times 10^{-10}$\,erg\,cm$^{-2}$\,s$^{-1}$ (integrated in the 0.1--100\,GeV energy range). This value agrees with the upper limit found by \cite{Wang2019}. By extrapolating the LAT light curve of Fig.~\ref{fig:lc_lat} at $t \sim 10.5$ h, we predict a 0.1--100\,GeV flux  of $\sim 5 \times 10^{-13}$\,erg\,cm$^{-2}$\,s$^{-1}$ if the flux decays $\propto t^{-2.2}$ (as found in the first 500 seconds). Instead, if after $\sim 500$\,s  the flux temporal decay becames shallower (e.g. $\propto t^{-1}$, consistently with the X--ray data\footnote{\url{https://www.swift.ac.uk/xrt_curves/00848890/}}), the flux at 10.5 hours would be $\sim 8 \times 10^{-11}$ erg cm$^{-2}$ s$^{-1}$. 

Therefore, the true LAT flux at around 10\,h should be between $\sim 5 \times 10^{-13}$ \,erg\,cm$^{-2}$\,s$^{-1}$ and the upper limit $8.5\times 10^{-10}$  erg cm$^{-2}$ s$^{-1}$. We note if the flux decays  $\propto t^{-1}$, its value at 10.5 hours is similar to the one detected by H.E.S.S. at higher energies. 
In conclusion, the LAT data are insufficient to unveil the nature of the emission component detected by H.E.S.S..

\section{Summary and Conclusions}\label{sec:conclusions}

In this paper we have studied both the prompt and the very early afterglow emission of GRB~180720B. 
The spectral evolution of the burst up to  500\,s after the trigger time is shown in Fig.~\ref{fig:sed}. 
The SED evolution (Fig.~\ref{fig:sed}) shows the emergence of the afterglow component in the LAT energy range and the progressive turning off of the prompt emission (which dominates up to 70 - SED I and II -  in the 0.01--1 MeV energy range). 

The LAT light curve shows a peak which we interpret as due to the deceleration of the outflow (Fig.~\ref{fig:lc_lat}). The peak time provides an estimate of the bulk Lorentz factor $\Gamma_0 \sim 300~(150)$ for a homogeneous (wind-like) circumburst medium.

The SED of the first three time intervals were fitted (through XSPEC) with a physical model of synchrotron emission from a relativistic population of injected electrons. 
From the SED accumulated over the first 35 sec, we find that the emitting electrons should be injected with a steep powerlaw energy distribution with $p \sim 4.8$. 

From the spectral features of the first 35\,s SED, we were able to estimate the intrinsic physical parameters of the emission region during the prompt phase. In particular, the comoving magnetic field $B' \sim 1$\,G, the minimum electron Lorentz factor $\gamma_{\rm min} \sim 10^6$ and the total number of electrons $N_e \sim 10^{51}$. Large values of $\gamma_{\rm min}$ and $p$ require an efficient acceleration mechanism and a "soft" energy distribution of accelerated particles which challenge current understanding of particle  acceleration through mildly relativistic shocks or magnetic reconnection. 

We also find that \fe/LAT did not detect any emission at the epoch (10.5 hours) of the claimed H.E.S.S. detection and derive an upper limit on the LAT flux (0.1--100 GeV) which is above the flux reported by H.E.S.S. (in the 100--440 GeV energy range). The extrapolation of the LAT light curve at 10.5 hours gives a 0.1--100 GeV flux consistent with the flux level detected by H.E.S.S.. LAT data therefore do not allow us to unveil the nature of the emission component detected $>$100 GeV. 

%==========================================================================
\begin{acknowledgements}
This research has made use of data obtained through the High Energy Astrophysics Science Archive Research Center Online Service, provided by the NASA/Goddard Space Flight Center, and specifically this work made use of public \fe/GBM data and of data supplied by the UK Swift Science Data Centre at the University of Leicester.
We also would like to thank for support the implementing agreement ASI-INAF n. 2017-14-H.0. We acknowledge the PRIN INAF CTA-SKA project ''Towards the SKA and CTA era: discovery, localisation, and physics of transient sources" and the PRIN-MIUR project "Figaro" for financial support.
\end{acknowledgements}

\bibliographystyle{aa} 
\bibliography{references}

\end{document}